\documentclass[english,aps,prl,showpacs,amsmath,amssymb,twocolumn]{revtex4}
\usepackage{graphicx}
\usepackage{bm}
\usepackage[latin1]{inputenc}
\begin{document}
\title{Single-slit diffraction and the Heisenberg principle for position and momentum}
\author{Winfrid G\"orlich, Ingo Hoffmann, Thomas Sch\"urmann}
\email{<th.schuermann@gmail.com>}
\affiliation{J\"ulich Supercomputing Centre, J\"ulich Research Centre, 52425 J\"ulich, Germany}

\begin{abstract}
Monochromatic light is prepared by a single slit of spatial width $\Delta x$. The diffraction pattern is considered to obtain the corresponding probability density of the momentum. Let $P$ be the probability weight to measure a momentum in a finite window of width $\Delta p$. For $\Delta p\to 0$ the asymptotic law $P\sim\frac{\Delta x\Delta p}{h}$ is verified, while $h$ is Planck's constant of action.\\
\\
\\
Keywords: Uncertainty principle; Measurement process; Planck's constant; Diffraction experiment; Quantum mechanics;
\end{abstract}

\pacs{03.65.-w, 03.65.Ta, 03.65.Db}
\maketitle

\section{Introduction}
The diffraction of a plane wave by a slit has often been discussed as an illustration of Heisenberg's uncertainty
relations and their role in the process of measurement. Let us begin with the ordinary case of a single particle passing
through a slit in a diaphragm of some experimental arrangement. Even if the momentum of the particle is completely known
before it impinges on the diaphragm, the diffraction of the plane wave by the slit will imply an uncertainty in the
momentum of the particle after it has passed the diaphragm, which is greater the narrower the slit.

Now the width of the slit, say $\Delta x$, may be taken as the uncertainty of the position of the particle relative to
the diaphragm, in a direction perpendicular to the slit. It is seen from de Broglie's relation between momentum and
wave-length that the uncertainty $\Delta p$ of the momentum of the particle in this direction is correlated to $\Delta x$
by means of Heisenberg's general principle $\Delta x\Delta p\sim h$. In his celebrated paper \cite{H27} published in
1927, Heisenberg attempted to establish this quantitative expression as the minimum amount of unavoidable momentum
disturbance caused by any position measurement.

Heisenberg himself did not give a unique definition for the 'uncertainties' $\Delta x$ and $\Delta p$, but estimated them
by some plausible measure, in each case separately. In his lecture \cite{H30} he emphasized his principle by the formal
refinement
\begin{eqnarray}\label{ungl}
\Delta x\Delta p\gtrsim h.
\end{eqnarray}

On the other hand, it was Kennard \cite{K27} in 1927 who first proved the now-popular inequality
\begin{eqnarray}\label{K}
\sigma_x\sigma_p\geq \hbar/2
\end{eqnarray}
with $\hbar=h/2\pi$, and $\sigma_x$, $\sigma_p$ being the ordinary standard deviations of position and momentum.
Heisenberg himself proved relation (\ref{K}) for Gaussian states \cite{H27,H30}.\\

Clearly, the statistical dispersion principle (\ref{K}) and the common statement of the uncertainty principle
(\ref{ungl}) are not equivalent or even closely related \cite{B70}. The statement (\ref{ungl}) refers to errors of {\it
simultaneous} measurements of position and momentum on one system. The position and the momentum are both considered for
the same particle and the key observation is that the initial measurement of the position necessarily disturbs the
particle, so that the momentum is changed by the preparation. Actually, this is the situation when particles pass a
slit.

On the other hand, (\ref{K}) refers to statistical spreads in ensembles of measurements on similar prepared systems. But
only one of them, either the position or the momentum, is measured on any one system. So there is no question of one
measurement interfering with the other.\\

In what will follow, we just focus on {\it simultaneous} measurement processes corresponding to the uncertainty principle
(\ref{ungl}). Initially, we will discuss the precise definition of $\Delta x$ and $\Delta p$ and their meaning in terms
of the measurement process under consideration. As we will see in the following section, some refinements of the
statement (\ref{ungl}) seem necessary to obtain a well-defined experimental setup. Finally, we present a simple laser experiment and verify the uncertainty principle.

\section{The single-slit experiment}
In the single-slit diffraction experiment, a monochromatic plane wave, representing an incoming beam of particles with
momentum $p_0$, impinges on a wall that contains an infinitely long slit of width $\Delta x$. The diffracted particles
are observed on a screen placed at a distance $L$ behind the slit. Without loss of generality, the normalized wave function in position spaces within the slit will be considered to be
\begin{eqnarray}
\psi(x)=\left\{ \begin{array}{r@{\quad\quad}l}
\frac{1}{\sqrt{\Delta x}} & \text{if }\quad |x|\leq \Delta x/2\\
0 & \text{if }\quad|x|>\Delta x/2 \label{abl}
\end{array} \right.
\end{eqnarray}
where $x$ is the position coordinate in the direction perpendicular to the slit. The most natural measure of the
uncertainty in position is the width of the slit $\Delta x>0$, while any particle reaching the screen has to pass the
slit in advance. Let $p$ be the momentum component along the $x$-direction. Then, all particles of the sample acquire a
momentum spread on passing through the slit in accordance to the distribution
\begin{eqnarray}\label{psip}
|\varphi(p)|^2 =\frac{\Delta x}{h} \left[\frac{\sin(\frac{\pi\Delta x}{h}\,p)}{\frac{\pi\Delta x}{h}\,p}\right]^2.
\end{eqnarray}
In quantum mechanics, the latter is obtained by Fourier transform of the initial wave function (plane wave) reduced by the slit. \\

Although the position uncertainty $\Delta x$ is clearly defined by the width of the slit, the uncertainty $\Delta p$ of the momentum has many appearances. In the usual analysis it is evaluated as the width of the main peak in the diffraction
pattern at the screen. Typically, the latter is chosen as twice the value of the first interference minimum (FIM), or equal
to the full width at the half maximum (FWHM) \cite{S69}\cite{Le69}\cite{Z03}. The choice between both is sometimes forced
by practical purposes since a high quality diffraction pattern is hard to obtain. For instance, in the
case of heavy massive particles the minima of the diffraction pattern do not always reach zero at their minima.
However, such measures are mostly based on the probability weight of the momentum density inside the width $\Delta p$
around the main peak.

The probability of detecting particles with momentum inside the interval $\Delta p$ is formally given by integrating
the momentum density of the particles. In our example, we obtain the following expression:
\begin{eqnarray}\label{Prob1}
P(\Delta p|\Delta x;\varphi)=\int_{-\Delta p/2}^{\Delta p/2}|\varphi(p)|^2\,dp
\end{eqnarray}

Actually, this probability is a conditional probability and dependent on both measurement precisions $\Delta x$ and $\Delta p$, ensuring the tradeoff between the complimentary observables.
\begin{figure}[ht]
\includegraphics[width=9.0cm,height=7.0cm]{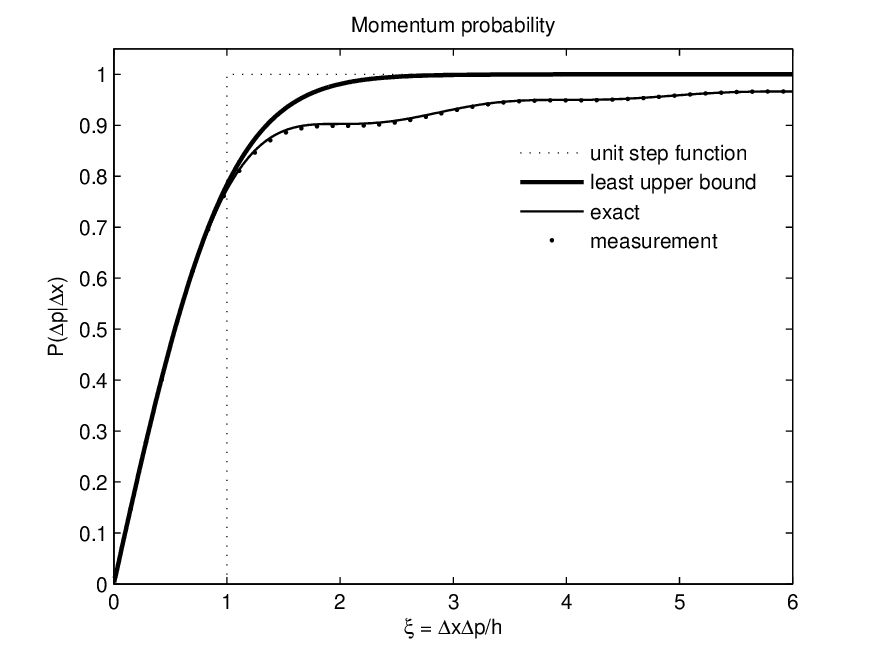}
\caption{The Heisenberg inequality (\ref{ungl}) is schematically expressed by a step-function at $\xi=1$. For the plane wave
(exact) the conditional probability (\ref{Prob1}) is monotonic increasing and approaches $\sim\xi$ for $\xi \to 0$. The least upper bound is also shown (see text).} \label{fig1}
\end{figure}

In {Fig.\,\ref{fig1}}, we see the measurement probability for a plane wave with respect to the parameter 
\begin{eqnarray}\label{xii}
\xi=\frac{\Delta x\Delta p}{h}.
\end{eqnarray}

The FWHM and the FIM then correspond to the special cases $\xi=0.89$ and $\xi=2$ respectively. The Heisenberg inequality (\ref{ungl}) is {\it schematically} expressed by the 'step-function' at $\xi=1$.

On the other hand, there is a general least upper bound of the probability (\ref{Prob1}) which is {\it independent} of the wave function. This bound was first obtained by Slepian, Landau and Pollak in the study of bandlimited functions for time-frequency analysis in signal-theory \cite{SP61}\cite{LP61}\cite{S65}. Its interpretation for position and momentum observables can be found in \cite{L72}\cite{L86}\cite{TS08}. In the quantum limes $\xi\to 0$ this bound approaches zero of order $\sim \xi$ (see Fig. \ref{fig1}).

In the following section, we will consider a simple (textbook) setup for an experimental verification of the momentum probability (\ref{Prob1}) applied to the particular case of a monochromatic wave.

\section{Experimental setup}\label{exp}
The diffraction pattern associated with the probability density (\ref{psip}) is obtained on a distant screen. That is, for large distances $L$ the momentum substitution $p\to x p_0/L$ is applied, while $x$ is the coordinate perpendicular to the beam measured on the screen and $p_0=h/\lambda_0$ is the initial mean momentum of the particles.

The incident wave may represent either the Schr\"{o}dinger wave function of a particle of mass $m$, in which case the angular frequency is $\omega=\hbar k^2/2m$, or else it may represent the electric field amplitude of an electromagnetic wave, in which case $\omega=ck$. The latter approach is represented by the Helmholtz equation and the mathematical formulation is similar in both cases. By simple algebraic substitution we obtain the following intensity pattern at the screen in terms of $x$ in the Fraunhofer
approach
\begin{eqnarray}\label{psix}
|\tilde\varphi(x)|^2 =\frac{\Delta x}{\lambda_0 L} \left[\frac{\sin(\frac{\pi \Delta x}{\lambda_0}\,x)}{\frac{\pi
\Delta x}{\lambda_0}\,x}\right]^2
\end{eqnarray}
and the screen position $x$ is related to the parameter
\begin{eqnarray}\label{xi_exp}
\xi = \frac{2 \Delta x}{\lambda_0 L}\,x.
\end{eqnarray}

The latter is the key quantity relating the theoretical predictions of {Fig.\,\ref{fig1}} within the slit to the
intensity pattern at the screen.
It is known from Fourier-Optics \cite{Go68} that a thin converging lens with focal length $f$ performs a Fourier transformation between the front and rear focal plane. Furthermore, it reduces the limit $L \to \infty$ to the finite length $L = f$, and equation (\ref{psix}) holds.
\begin{figure}[ht]
\scalebox{0.34}{\includegraphics[7.0cm,0.0cm][25.0cm,12cm]{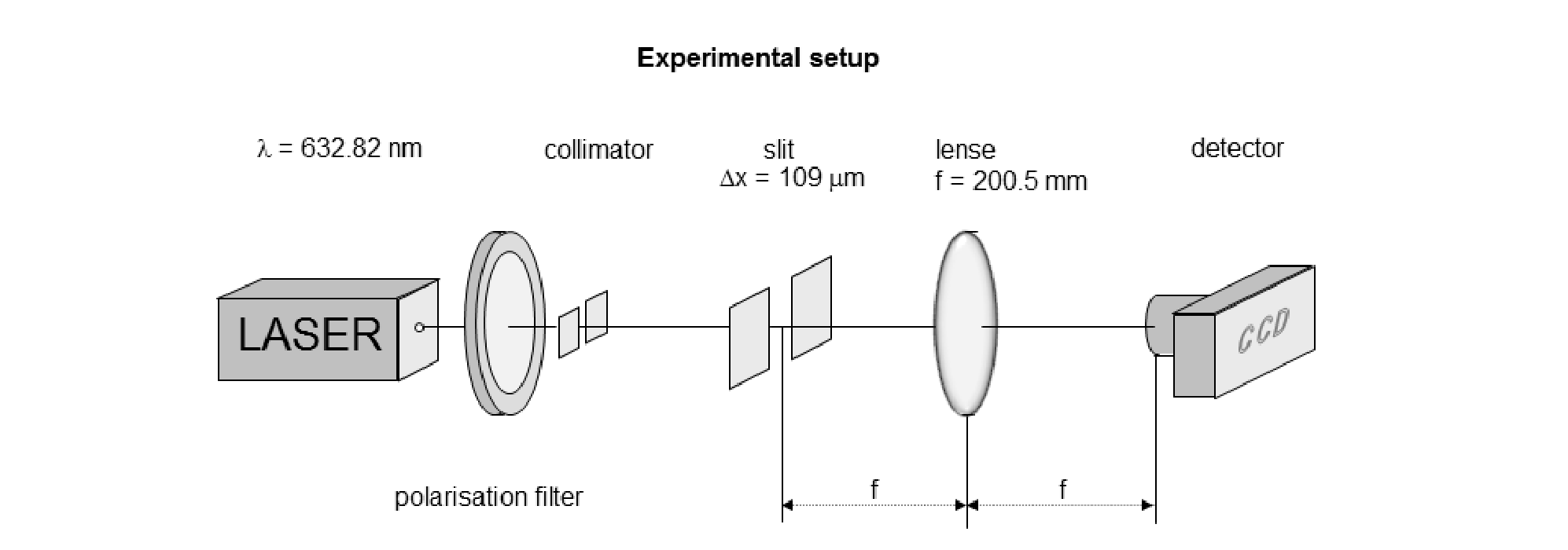}}
\caption{The intensity of the diffraction pattern is measured by a CCD-Sensor. The width of the slit is $\Delta x = 109 \pm 2\ \mu$m and the focal length of the lens is $f = 200.5$ mm. }\label{fig2}
\end{figure}

We apply laser light of the spontaneous and stimulated emission between the $3s^2$ and $2p^4$ states which results in a
wavelength of $\lambda_0=\text{632.82\ n}$m, the typical operating wavelength of a HeNe-Laser (power: $\text{0.5 m}$W). The
intensity of the beam is controlled by a polarization filter and subsequently collimated (see {Fig.\,\ref{fig2}}). The fixed width of the slit is $\Delta x = 109 \pm 2\ \mu$m and the focal length of the thin lens is $f = 200.5$ mm.

The intensity of the diffraction pattern is measured by a Charge-Coupled-Device Sensor (CCD-Sensor). It is located at
the distance of $2f$ behind the slit. The measurement range and the resolution of the CCD-Sensor is given
by 3648 pixels and the size of a single pixel $8\ \mu $m. The exposure time for a single snapshot is $1$ ms.

\begin{figure}[ht]
\includegraphics[width=9.0cm,height=6.5cm]{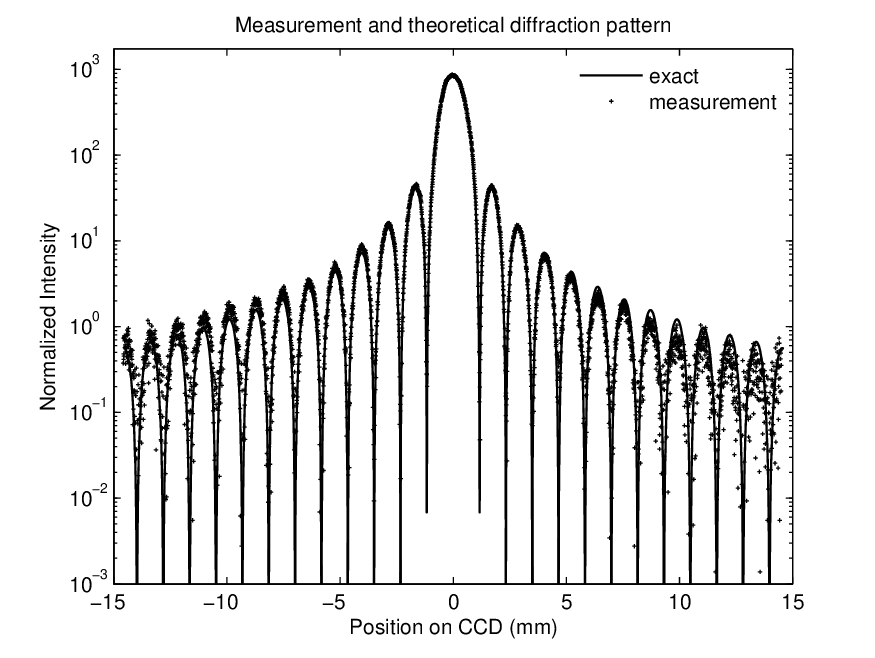}
\caption{Measured versus theoretical probability density in semi-logarithmic representation for 12 orders of diffraction. The theoretical prediction fits the data for about 4 orders of magnitude.  } \label{fig3}
\end{figure}

In {Fig.\,\ref{fig3}}, we see the average diffraction pattern of the measurement compared to the exact density
corresponding to (\ref{psix}) in semi-logarithmic scales. The error bars are very small and thus not expressed in the figure. The experimental data fit very well to the exact computation up to 12 orders of diffraction. By numerical integration of the data we obtain the proabability (\ref{Prob1}).

It should be mentioned that the main peak of the measured histogram is not normalized by the maximum of the
central peak. Instead, we compute the total voltage over all pixels and the partial area of the momentum interval $\Delta p$ of the empirical density is obtained by numerical integration. The monotonic increasing behavior of the distribution is shown in {Fig.\,\ref{fig1}}.  All measurement results fit very well to the theoretical computation of (\ref{Prob1}) over the whole range $\xi>0$.

\section{Summary}

It has been shown that there are measurement events beyond the inequality $\Delta x\Delta p\geq h$. Moreover, we verified that the probability weight of such events approach zero by the scaling law $P\sim\frac{\Delta x\Delta p}{h}$, when $\Delta x$ or $\Delta p$ is sufficiently small. Finally, it has been verified that the measurement intensity satisfies the least upper bound predicted in literature.

\acknowledgments

\newpage{}

\begin{thebibliography}{99}
\bibitem{H27} W. Heisenberg, Z. Phys. {\bf 43}, 172 (1927).
\bibitem{H30} W. Heisenberg, \emph{The Physical Principles of the Quantum Theory}, (University of Chicago Press, Chicago,
    1930) [Reprinted by Dover, New York (1949, 1967)].
\bibitem{K27} E. H. Kennard, Z. Phys. {\bf 44}, 326 (1927).
\bibitem{B70} L. E. Ballentine, Rev. Mod. Phys. {\bf 42}, 358 (1970).
\bibitem{S69} C. G. Shull, Phys. Rev. {\bf 179}, 752 (1969).
\bibitem{Le69} J. A. Leavit, F. A. Bills, Am. J. Phys. {\bf 37} (9), 905 (1969).
\bibitem{Z03} O. Nairz, M. Arndt and A. Zeilinger, Phys. Rev. A {\bf 65}, 032109 (2002);\\
\text{http://arxiv.org/abs/quant-ph/0105061v1}.
\bibitem{SP61} D. Slepian and H. O. Pollak, Bell Syst. Techn. J. {\bf 40}, 43 (1961).
\bibitem{LP61} H. J. Landau and H. O. Pollak, Bell Syst. Techn. J. {\bf 40}, 65 (1961).
\bibitem{S65} D. Slepian , J. Math. Phys. {\bf 44}, 99 (1965).
\bibitem{L72} A. Lenard, J. Functional Anal. {\bf 10}, 410 (1972).
\bibitem{L86} P. J. Lahti, Rep. Math. Phys. {\bf 23}, 289 (1986).
\bibitem{TS08} T. Sch\"urmann, Acta Phys. Pol. {\bf B 39}, 587 (2008);
    \text{http://th-www.if.uj.edu.pl/acta/vol39/pdf/v39p0587.pdf}.
\bibitem{Go68} J. W. Goodman, \emph{Introduction to Fourier Optics}, {McGraw-Hill Inc., San Francisco, 1968}
\end{thebibliography}
\end{document}